\documentclass{article}
\usepackage{multicol}
\usepackage[margin=2cm]{geometry}
\usepackage{titling}
\usepackage{authblk}
\usepackage[superscript,biblabel]{cite}
\usepackage{graphicx}
\usepackage{siunitx}
\usepackage{textcomp}
\usepackage[version=4]{mhchem}
\usepackage{float}
\usepackage{textgreek}
\begin{document}

\title{Chemical Strain Engineering of MAPbI$_3$ Perovskite Films}
\author[1,2]{Yenal Yalcinkaya}
\author[1]{Ilka M. Hermes}
\author[3]{Tobias Seewald}
\author[1]{Katrin Amann-Winkel}
\author[1]{Lothar Veith}
\author[3]{Lukas Schmidt-Mende}
\author[1,2]{Stefan A.L. Weber}
\affil[1]{Max Planck Institute for Polymer Research, Ackermannweg 10, 55128 Mainz, Germany}
\affil[2]{Institute of Physics, Johannes Gutenberg University Mainz, Duesbergweg 10-14, 55128 Mainz, Germany.}
\affil[3]{Department of Physics, University of Konstanz, Universitätsstr. 10, 78464, Germany}
\setcounter{Maxaffil}{0}
\renewcommand\Affilfont{\itshape\small}

\maketitle
\begin{abstract}

This study introduces a new chemical method for controlling the strain in methylammonium lead iodide (MAPbI$_3$) perovskite crystals by varying the ratio of Pb(Ac)$_2$ and PbCl$_2$ in the precursor solution. To observe the effect on crystal strain, a combination of piezoresponse force microscopy (PFM) and X-ray diffraction (XRD) is used. The PFM images show an increase in the average size of ferroelastic twin domains upon increasing the PbCl$_2$ content, indicating an increase in crystal strain. The XRD spectra support this observation with strong crystal twinning features that appear in the spectra. This behaviour is caused by a strain gradient during the crystallization due to different evaporation rates of methylammonium acetate and methylammonium chloride as revealed by time-of-flight secondary ion mass spectroscopy (ToF-SIMS) and grazing incidince x-ray diffraction (GIXRD) measurements. Additional time-resolved photoluminescence (TRPL) show an increased carrier lifetime in the MAPbI$_3$ films prepared with higher PbCl$_2$ content, suggesting a decreased trap density in films with larger twin domain structures. The results demonstrate the potential of chemical strain engineering as an easy method for controlling strain-related effects in lead halide perovskites.

\end{abstract}

\clearpage
  
\section{Introduction}
    
    Hybrid lead halide perovskites \cite{kojima2009organometal,lee2012efficient,kim2012lead,yoo2019interface,jaramillo2015bright,ashar2018sensors} show remarkable properties such as a direct adjustable band gap \cite{protesescu2015nanocrystals,stoumpos2013semiconducting,jacobsson2016exploration}, high defect tolerance \cite{steirer2016defect}, and long charge carrier lifetimes \cite{johnston2016hybrid,stranks2013electron,shi2015low} that make them ideally suited as absorber materials for photovoltaic applications. An interesting effect that has been observed in many hybrid lead halide perovskite materials is that the optoelectronic properties can be strongly influenced by external or internal strain. Thus, strain engineering can be a useful method to control and tune the optoelectronic properties or structural stability of perovskite materials\cite{moloney2020strain,yuce2021understanding}. For example, Zhu et al. showed improved charge extraction at the perovskite-hole transport layer interface by eliminating the strain gradient, flattening the valence band\cite{zhu2019strain}. Kim et. al. used strain engineering to relax the crystal lattice of formamidinium lead iodide (FAPbI$_3$), enabling improved solar cell performance\cite{kim2020impact}. In addition, strain engineering had also been used by various groups to increase the stability of perovskite structures. Zhao et al. showed strain engineering of MAPbI$_3$ by mechanichally bending the substrate increased the stability\cite{zhao2017strained}. Strain engineering was also used to stabilize meta-stable phases such as CsPbI$_3$ and FAPbI$_3$ by introduction of biaxial strain via cooling a perovskite crystal clamped to a ceramic substrate \cite{steele2019thermal} or applying strain by heteroepitaxially growing the perovskite on another perovskite material \cite{chen2020strain}.
    
    \
    
    Another important strain-related effect observed in  methylammonium lead iodide (MAPbI$_3$) crystals is the formation of sub-granular ferroelastic twin domains. Ferroelastic twin domains in tetragonal MAPbI$_3$ were first observed via piezoresponse force microscopy (PFM) by our group \cite{hermes2016ferroelastic}, followed by many others\cite{huang2018ferroic,rohm2017ferroelectric,rothmann2017direct,liu2018chemical}. The domains form due to the strain resulting from the cubic to tetragonal phase change at 57\textdegree C.  Strelcov et al. observed via polarized optical microscopy (POM) that mechanical stress leads to a rearrangement of the twin domains in MAPbI$_3$, supporting the conclusion that the domains are ferroelastic \cite{strelcov2017ch3nh3pbi3}. Since our first report in 2016, ferroelastic twins have been observed by many other techniques, including transmission electron microscopy (TEM) \cite{rothmann2017direct}, X-ray diffraction (XRD)\cite{medjahed2020unraveling}, and neutron scattering\cite{breternitz2020twinning}.  
    
    \
    
    The existence of a sub-granular domain structure raises the question whether these domains have an influence on the electronic transport properties. Recently, Xiao et al. \cite{xiao2020benign} concluded from PL and lifetime mapping experiments that the ferroelastic twin domains are benign to recombination kinetics. To separate potential grain boundary effects, our group recently investigated large isolated MAPbI$_3$ grains. Using time-resolved PL microscopy, we found an anisotropic charge transport that was correlated to the ferroelastic twin domain structure in the crystal \cite{hermes2020anisotropic}. In particular, the charge diffusion perpendicular to the ferroelastic domains was slower compared to the charge diffusion parallel to the domains. 
    
    \
    
    Such an anisotropic diffusion would be interesting for guiding charge carriers, e.g. to the electrodes of an optoelectronic device. Thus, a targeted manipulation or engineering of the twin domain structure would be desirable. Next to domain manipulation through mechanical straining\cite{strelcov2017ch3nh3pbi3}, we showed that the ferroelastic twin domains structure changes when using different cooling rates after annealing \cite{hermes2020anisotropic,vorpahl2018orientation}. Here, a slower cooling rate resulted in a more dense and ordered domain structure whereas a higher cooling rate lead to disordered and less dense twin domains. Lastly, Röhm et al. demonstrated the domain manipulation via a lateral electric field \cite{rohm2020ferroelectric}. However, all the domain manipulation methods mentioned above are not easy to scale up for mass production.
    
    \
    
    In this study, we introduce a new method for controlling the mechanical strain during the crystal growth via the composition of the precursor solution. In particular, we use different ratios of Pb(Ac)$_2$ and PbCl$_2$ to synthesize MAPbI$_3$ thin films on glass substrates in order to adjust the overall crystallization of MAPbI$_3$ films. Acetate and chloride anions were chosen because they require similar annealing temperatures for evaporation/sublimation. Using a combination of PFM and XRD, we observed an increase in crystal strain connected to an increase in the average twin domain size upon increasing the PbCl$_2$ content. Additionally, we observed a change from stripe-like to rectangular domains, an effect that has been connected to increased crystal strain\cite{strelcov2017ch3nh3pbi3}. Based on time-of-flight secondary ion mass spectroscopy (ToF-SIMS) experiments where we found an accumulation of chloride at the substrate interface, we suggest that the strain during crystal growth is caused by the combined effect of low temperature crystallization and ion exchange during annealing. This chloride accumulation is most likely caused by different evaporation rate of the methylammonium acetate and sublimation rate of methylammonium chloride. Additional time-resolved photoluminescence (TRPL) showed an increased carrier lifetime in the MAPbI$_3$ films prepared with higher PbCl$_2$ content, suggesting a decreased trap density in films with larger twin domain structures. 

\section{Results and Discussion}

\subsection{Piezoresponse Force Microscopy (PFM)}

\begin{figure}[h!]
  \includegraphics[width=\linewidth]{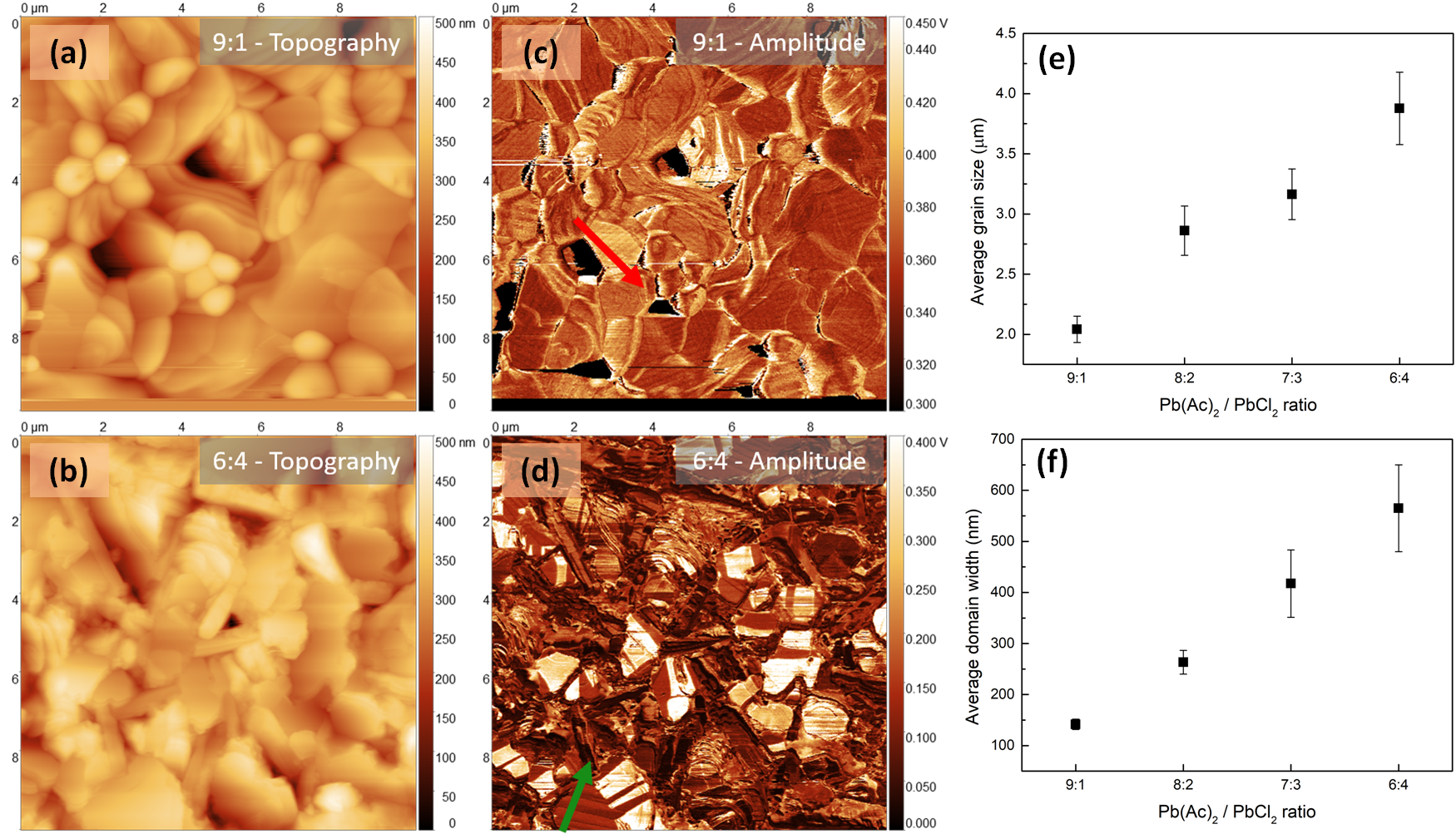}
  \caption{Topography PFM amplitude images of MAPbI$_3$ thin films with Pb(Ac)$_2$ / PbCl$_2$ ratios of (a) 9:1, and (b) 6:4. (c) Average grain size and domain width profiles of MAPbI$_3$ thin films with different Pb(Ac)$_2$ / PbCl$_2$ ratios.}
  \label{fig:PFM}
\end{figure}

We prepared several batches of MAPbI$_3$ films on glass with precuror ratios between 9:1 and 6:4  and investigated the topography and the lateral piezoresponse using AFM/PFM  (Figure \ref{fig:PFM}a-d and S2, details on the working mechanism of PFM are given in the methods section). Figure \ref{fig:PFM}e shows the average grain sizes obtained from several areas of the films with changing Pb(Ac)$_2$ / PbCl$_2$ ratios. The most obvious effect of the the precursor ratio change is an increase in average grain size ($<$d$>$) from 2.04\textpm 0.11 µm for the 9:1 Pb(Ac)$_2$ / PbCl$_2$ ratio to 3.9\textpm 0.3 µm for the 6:4 Pb(Ac)$_2$ / PbCl$_2$ ratio (Figure \ref{fig:PFM}e). A MAPbI$_3$ film prepared from pure Pb(Ac)$_2$ precursor had an even smaller average grain size of 0.85\textpm 0.31 µm (Figure S1, Supporting Information). The decrease in grain size upon lowering the PbCl$_2$ content comes from the decreased Cl or MACl amount which speeds up the crystallization of perovskite and leads to smaller grains\cite{zhang2015ultrasmooth}. Furthermore, the combined effect of MACl and DMF annealing seems to be causing a significant increase in grain size compared to MAPbI$_3$ film prepared from pure Pb(Ac)$_2$ precursor. In addition to the grain size, the general grain morphology also changed upon decreasing Pb(Ac)$_2$ / PbCl$_2$ ratio. In between the perovskite grains we observed more pinholes and undefined structures (Figure 1b). On average, the density of pinholes and undefined structures increases with decreasing Pb(Ac)$_2$ / PbCl$_2$ ratio (Figure S2, Supporting Information), suggesting these structures could be caused by residual PbI$_2$ or MAPbCl$_3$ that forms after the spin coating \cite{medjahed2020unraveling}. The reason for this effect could again be a slower crystallization in the MAPbI$_3$ film due to the lower volatility of MACl compared to MA(Ac). Later on, we will show that this interpretation is also supported by ToF-SIMS (Figure \ref{fig:vertical ToF-SIMS}),UV-Vis absorption (Figure S11, Supporting Information), and XRD reflection results (Figure \ref{fig:XRD-All}).

\

The lateral PFM amplitude images show the familiar striped ferroelastic domain structure \cite{hermes2016ferroelastic} (Figure \ref{fig:PFM}c and d). Compared to MAPbI$_3$ films with 9:1 Pb(Ac)$_2$ / PbCl$_2$ ratio (Figure \ref{fig:PFM}c), films prepared from the 6:4 precursor ratio show a less dense domain structure (Figure \ref{fig:PFM}d). In particular, we observed an increase in the width of the ferroelastic twin domains with increasing PbCl$_2$ content. Figure \ref{fig:PFM}f shows average ferroelastic twin domain width values gathered from various areas of the samples. The distances between the high amplitude and low amplitude areas revealed an average domain widths ($<$w$>$) of 137.5\textpm 10 nm for the MAPbI$_3$ film with 9:1 and 558.33\textpm 90 nm and 6:4 Pb(Ac)$_2$ / PbCl$_2$ ratio, respectively (Figure \ref{fig:PFM}f).

\

As twin domains are purely strain related, it seems obvious from the changing domain pattern that the strain in the films is influenced by different Pb(Ac)$_2$ / PbCl$_2$ ratios. The work by Strelcov et al. \cite{strelcov2017ch3nh3pbi3} has demonstrated that external stress can alter the domain structure. Their results show that upon increasing the strain within the MAPbI$_3$ films the stripe-shaped domains evolve into larger areas. Furthermore, the emergence of non-90\textdegree domain angles were observed upon increased external stress\cite{strelcov2017ch3nh3pbi3}. We also observed such domain shapes (Figure \ref{fig:PFM}), suggesting that the precursor mixing with Pb(Ac)$_2$ / PbCl$_2$ or MA(Ac) and MACl increases the strain within the Pb(Ac)$_2$ / PbCl$_2$ structure.

\subsection{X-Ray Diffraction}

\begin{figure}[h!]
  \includegraphics[width=\linewidth]{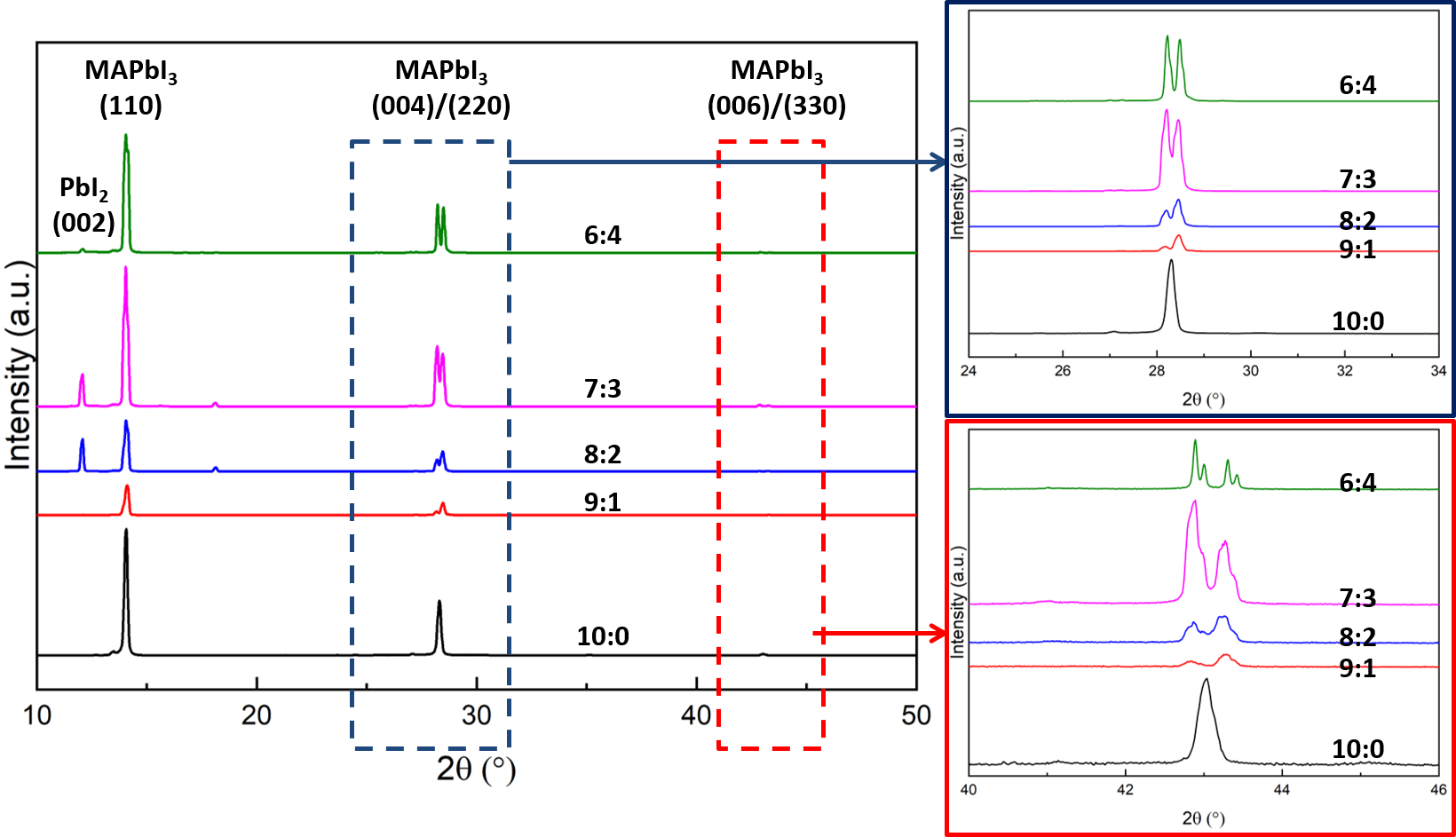}
  \caption{XRD patterns of MAPbI$_3$ thin films with Pb(Ac)$_2$ / PbCl$_2$ ratios of 10:0, 9:1, 8:2, 7:3 and 6:4. Intensities were shifted for clarity, the data are not normalized.}
  \label{fig:XRD-All}
\end{figure}

To support the hypothesis of precursor-induced strain adjustment in our MAPbI$_3$ films with different Pb(Ac)$_2$ / PbCl$_2$ ratios, we performed X-ray diffraction (XRD) measurements. In the XRD, all mixed precursor samples showed the typical reflections for MAPbI$_3$ with addition of PbI$_2$ (Figure \ref{fig:XRD-All} and S3) and some unidentified weak peaks around 18\textdegree{} that may belong to intermediates or a Lewis acid-base adduct due to retarded crystallization caused by increased PbCl2 \cite{lee2019precursor}. The MAPbI$_3$ (110) / PbI$_2$ (002) intensity ratio reaches to a maximum to some degree PbI$_2$ (002) peak is quenched by MAPbI$_3$ (110) intensity. We also observed a peak splitting around 28\textdegree{} due to \{00l\}/\{hk0\} twinning in  MAPbI$_3$ films \cite{medjahed2020unraveling} (Figure \ref{fig:XRD-All}, blue box) in MAPbI$_3$ films with Pb(Ac)$_2$ / PbCl$_2$. The intensity ratio of the (004) peaks to the (220) peaks around 28\textdegree{} increased when the Pb(Ac)$_2$ / PbCl$_2$ ratio was decreased. The lower inset of Figure \ref{fig:XRD-All} (red box) shows the XRD signal in the range between 40 and 46\textdegree{}. Again, as the Pb(Ac)$_2$ / PbCl$_2$ ratio was decreased, the (006)/(330) peak intensity ratios increased. Interestingly, we found a further splitting of the split peaks in samples prepared from a 6:4 Pb(Ac)$_2$ / PbCl$_2$ ratio, resulting in four peaks around 43\textdegree{}. Based on the work by Dang et al. \cite{dang2015bulk}, these additional peaks may belong to the (134) and (402) planes. Such further peak splitting in the XRD could be related to lower symmetry for smaller Pb(Ac)$_2$ / PbCl$_2$ ratios as a result from the formation of non-90\textdegree{} domains. A similar observation in XRD pattern of MAPbI$_3$ was observed by Leonhard et al.\cite{leonhard2021evolution} where the emerging peaks disappear upon further annealing, indicating strain relaxation. These results suggest that the larger domain areas in the PFM images in Figure \ref{fig:PFM}(b) belong to \{00l\} facets and that their XRD intensity increases due to increased area. 
Furthermore, the observation of a second peak splitting may indicate an increased residual stress resulted by different Pb(Ac)$_2$ / PbCl$_2$. 

\

To learn more about the homogeneous strain behaviour within the MAPbI$_3$ films, we analyzed the XRD peak shifts as a function of Pb(Ac)$_2$ / PbCl$_2$ ratio. The position for MAPbI$_3$ (110) peak for 10:0, 9:1, 8:2, 7:3, and 6:4 Pb(Ac)$_2$ / PbCl$_2$ ratios were observed at around 14\textdegree{}. One important observation in  MAPbI$_3$ (110) plane for 6:4 Pb(Ac)$_2$ / PbCl$_2$ ratio is the emerging shoulder at 14.14\textdegree{} (Figure \ref{fig:XRD-All}). This splitting of the (110) plane suggests that another twinning starts to take place in these samples, likely caused by a higher strain. The peak splitting trends of (004)/(220) and (006)/(330) suggest that the peak at the higher scattering angles would belong to \{hk0\} planes where the peak at the lower scattering angles would belong to the \{00l\} planes, possibly (002).  Furthermore, the positions of split MAPbI$_3$ (004)/(220) peaks were observed at 28.32\textdegree{} (only (220)), 28.18/28.47\textdegree{}, 28.2/28.46\textdegree{}, 28.21/28.45\textdegree{}, and 28.23/28.49\textdegree{} for 10:0, 9:1, 8:2, 7:3, and 6:4 Pb(Ac)$_2$ / PbCl$_2$ ratios, respectively. It is safe to assume that strain controls the trends observed in the XRD measurements since the enlarged domains were also observed in PFM as a result of increased strain. Here, a shift to lower scattering angles in MAPbI$_3$ (110) plane for Pb(Ac)$_2$ / PbCl$_2$ ratios lower than 9:1 which may indicate the emergence of another peak from the (002) plane. Alternatively, this shift may indicate lattice expansion based on the Bragg's law and interplanar distance formulas (Equation S1 and S2, Supporting Information). Moreover, a shift to the higher scattering angles for MAPbI$_3$ (004) plane and a shift to lower scattering angles for MAPbI$_3$ (220) was observed with decreasing Pb(Ac)$_2$ / PbCl$_2$ ratio which may indicate lattice shrinking and lattice expansion, respectively (Equation S1 and S2, Supporting Information). 

\

Lastly, we have also performed an XRD measurement on a MAPbI$_3$ film made from a precursor solution containing only MAI and Pb(Ac)$_2$. This sample in particular was not annealed but only dried at room temperature in the glovebox since MA(Ac) is much more volatile compared to MACl and MAPbI$_3$ can form without any heat treatment. 
The resulting films had a very small grain structure, which made PFM experiments impossible. Therefore, the twinning was only monitored via XRD for this sample (Figure S4a, Supporting Information). The resulting XRD pattern exhibited larger peak splitting, even more distinct then MAPbI$_3$ with 6:4 Pb(Ac)$_2$ / PbCl$_2$ ratio. This result supports the notion of room temperature twinning in MAPbI$_3$ \cite{breternitz2020twinning}. Also, the stronger (004) signal in non-annealed MAPbI$_3$ film made from only MAI and Pb(Ac)$_2$ compared to  annealed samples in Figure \ref{fig:XRD-All} is because of strain relaxation upon annealing.

\subsection{Chemical Gradients and Strain Mechanism}

To synthesize the MAPbI$_3$, methylammonium iodide (MAI), lead acetate (Pb(Ac)$_2$), and lead chloride (PbCl$_2$) precursors were utilized. In this synthesis, the molar ratio of MA precursor to Pb precursor is 3:1 as can given in Eq (1):

\begin{center}
    \ce{3 MAI + PbX_2 -> MAPbI_3 + 2MAX}    
    \label{eq:synthesis_reaction}
\end{center}

where X can be Ac, Cl, or I. The MAX compound is volatile and evaporates during the annealing of the film. The volatility and thus the evaporation rate of MAX depends on the X anion used in the precursor solution. Therefore, the crystallization kinetics in the MAPbI$_3$ film is controlled by the evaporation rate of the MAX compound. 

\begin{figure}[H]
    \begin{center}
       \includegraphics[width=0.5\linewidth]{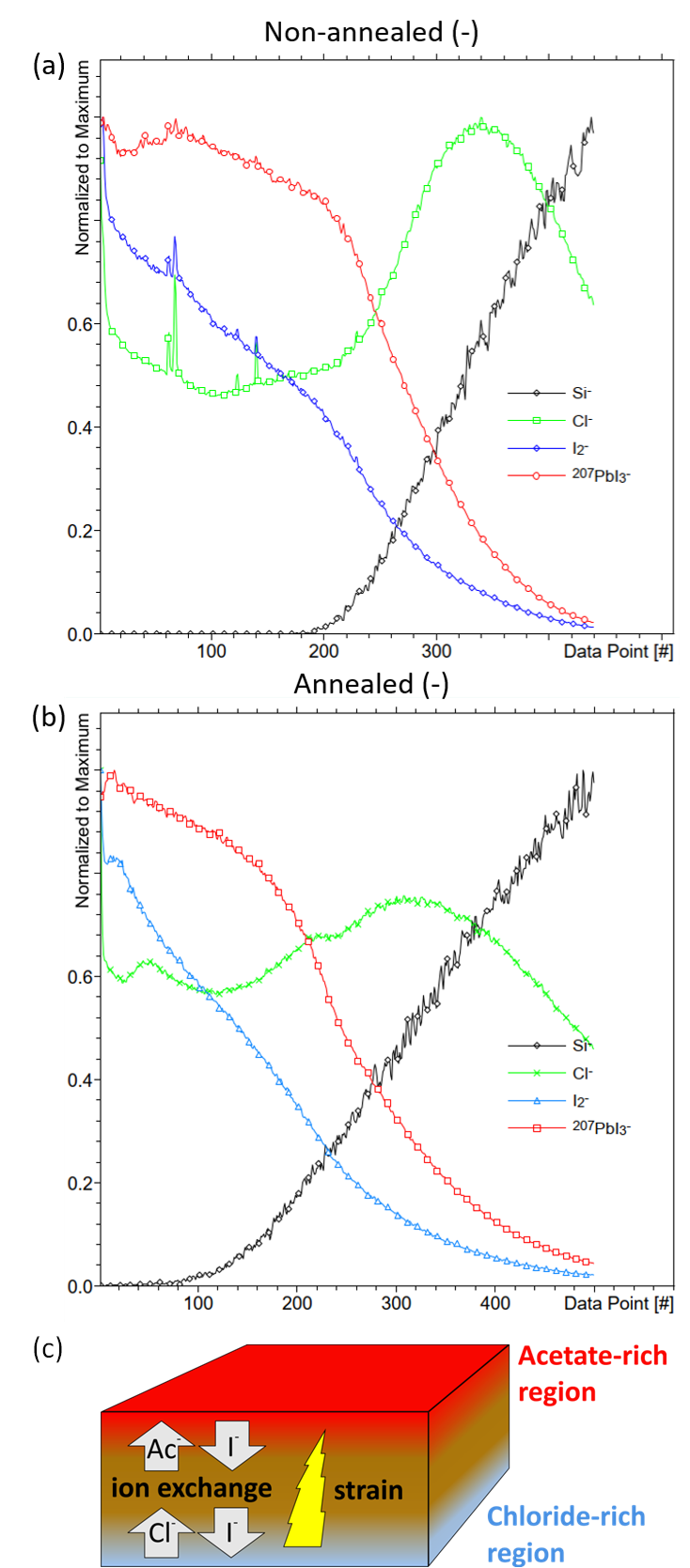}
       \caption{Negative chemical depth profiles of (a) non-annealed and (b) annealed MAPbI$_3$ film made with 6:4 Pb(Ac)$_2$ / PbCl$_2$ ratio. (c) An illustration of ion exchange reactions.}
       \label{fig:vertical ToF-SIMS}
    \end{center}
\end{figure}

The observed changes in the ferroelastic twin domain structure in PFM and the peak splitting behaviour in the XRD spectra suggest a correlation between the strain in the MAPbI$_3$ films and the Pb(Ac)$_2$ / PbCl$_2$ ratio in the precursor solution. A recent study by Medjahed et al. \cite{medjahed2020unraveling} has investigated the strain in MAPbI$_3$ films prepared from a 3MAI/PbCl$_2$ solution by monitoring the twinning via XRD. According to this study, MAPbCl$_3$ forms within the film prior to annealing and I$^-$ / Cl$^-$ ion exchange occurs during annealing. This process gives rise to strain formation within the MAPbI$_3$ film and eventually results in twin formation. 

\

To investigate chemical kinetics within the perovskite film, we carried out ToF-SIMS. The profiles for both PbI$_3^-$ and Cl$^-$ show a pronounced chemical gradient within the film (Figure \ref{fig:vertical ToF-SIMS}a and b). In the non-annealed film, we observed elevated levels of Cl$^-$ at the bottom of the film compared to static/central part of the profile line. As suggested previously, there might be volatile compounds on the surface involving Cl$^-$ which could be the reason for high Cl$^-$ amount on the surface. High Cl$^-$ amount in the depths of the film on the other hand shows that Cl$^-$ ions tend to localize closer to the substrate in the mixed precursor MAPbI$_3$ film as they do in pure MAI/PbCl$_2$ films \cite{unger2014chloride}. A similar trend was also observed for Cl$^-$ in the annealed MAPbI$_3$ film as well, however with an increased uniformity through the film. This can be an indication of I$^-$ / Cl$^-$ ion exchange that leads to the formation of MAPbI$_3$. 

\

The depth profiles in both pristine and annealed MAPbI$_3$ films show I$_2^-$ and PbI$_3^-$ signals with the opposite trend of Cl$^-$. After the precursor deposition, MAPbI$_3$ grains form closer to the surface while Cl-rich species stay at the bottom of the film. Upon annealing, MACl reaches the surface through the ion exchange. This ion exchange reaction of MAPbCl$_3$ is expected give rise to a vertical strain gradient. An illustration of this vertical ion exchange reaction is given in Figure \ref{fig:vertical ToF-SIMS}c.

\begin{figure*}[h!]
    \begin{center}
       \includegraphics[width=\linewidth]{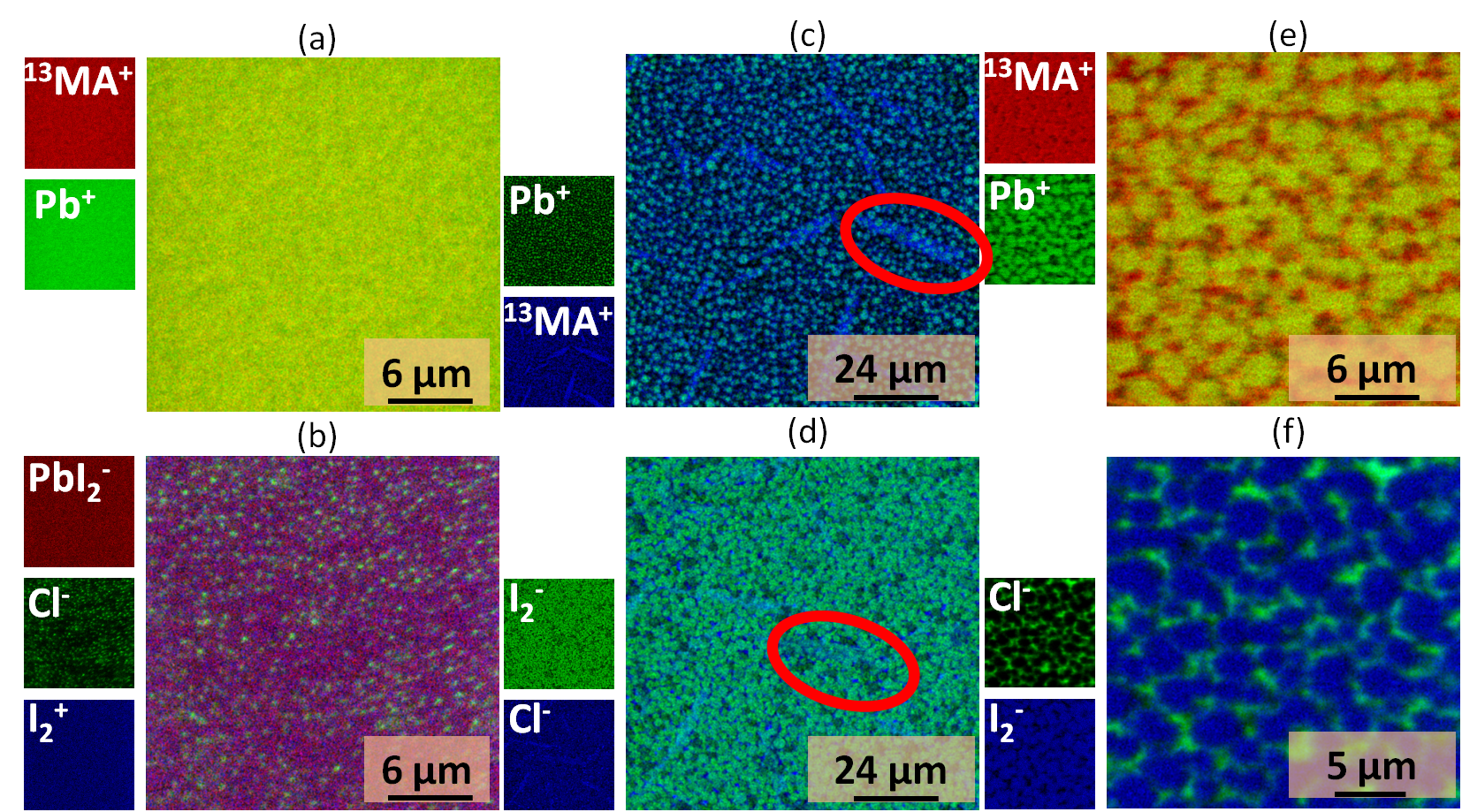}
       \caption{Lateral ion distributions in (a,b) non-annealed and (c,d,e,f) annealed MAPbI$_3$ films made from 6:4 Pb(Ac)$_2$ / PbCl$_2$ ratio.}
       \label{fig:ToF-SIMS_RGB}
    \end{center}
\end{figure*}

The depth profiles for positive ions (Figure S7) show a uniform Pb$^+$ ion distribution and a slight gradient in MA$^+$ ion distribution. The non-uniform distribution of MA$^+$ ion could also be contributing to the vertical strain gradient as observed before\cite{liu2020strain}. 

\

To investigate a potential strain gradient along the vertical direction of the film, we employed grazing incidence x-ray diffraction (GIXRD). GIXRD enables to change the x-ray penetration into the film by changing the incidence angle, providing depth-dependent crystal structure information. The GIXRD spectra of a MAPbI$_3$ film with 6:4 Pb(Ac)$_2$ / PbCl$_2$ ratio shows a change in XRD reflection around 14\textdegree (Figure S6, Supporting Information). Deconvolution of the signal reveals a clear shift in the lattice parameters as the x-ray penetration into the perovskite film changes. These shifts are a clear signature of a vertical strain gradient.

\

The lateral chemical distribution was also determined using ToF-SIMS on non-annealed and annealed MAPbI$_3$ films with 6:4 Pb(Ac)$_2$ / PbCl$_2$ ratio (Figure \ref{fig:ToF-SIMS_RGB}). The ion distribution in non-annealed MAPbI$_3$ film reveals uniform distribution of MA$^+$, Pb$^+$, and I$_2^-$ ions after spin coating whereas Cl$^-$ ions show a distribution pattern (Figure \ref{fig:ToF-SIMS_RGB}a and b). Based on the information we obtained from XRD measurements on non-annealed samples (Figure S5, Supporting Information), perovskite films at this stage should be consisting of MAPbI$_3$ and MAPbCl$_3$. Therefore, we suggest the Cl-rich areas in the film are mostly MAPbCl$_3$ areas where the acetate-rich areas quickly transform into MAPbI$_3$. Upon annealing, some morphological and chemical changes occured within the perovskite film: MA- and Cl-rich needle-like structures are prominent in the ToF-SIMS images (blue images in Figure \ref{fig:ToF-SIMS_RGB}c and d, marked with red), similar to the needle structures that we observed in the AFM images (Figure \ref{fig:PFM} b). In contrast, the Pb$^+$ distribution was much more homogeneous (Figure \ref{fig:ToF-SIMS_RGB}c), whereas the I$_2^-$ concentration was reduced in the needle-like structures (Figure \ref{fig:ToF-SIMS_RGB}d), suggesting these structures could be MACl that forms after I$^-$ / Cl$^-$ ion exchange. At larger magnification, (Figure \ref{fig:ToF-SIMS_RGB}e and d), we see that grain interiors are rich in MA$^+$, Pb$^+$, and I$_2^-$, whereas the exterior of the grains are rich in Cl$^-$. However, there are small areas where the MA$^+$ and Pb$^+$ maps do not exactly match. These map mismatches suggest that the structures outside of MAPbI$_3$ grains could be belonging to structures such as PbI$_2$ and MACl that is not sublimated yet. In addition, we observed small amounts of Cl$^-$ inside of the MAPbI$_3$ grains, suggesting a minor Cl$^-$ doping could be taking place during fabrication. The ion maps show that next to a vertical chemical gradient there is also a lateral chemical gradient within the perovskite films (Figure \ref{fig:vertical ToF-SIMS}). Since the correlation between the vertical chemical gradient and the strain gradient in the perovskite film is well established through our GIXRD results (Figure S6, Supporting Information) and supported by other studies \cite{medjahed2020unraveling,liu2020strain}, we conclude that lateral strain will be present within the films, as well.

\subsection{Optical Measurements}

\begin{figure}[ht]
  \includegraphics[width=0.8\linewidth]{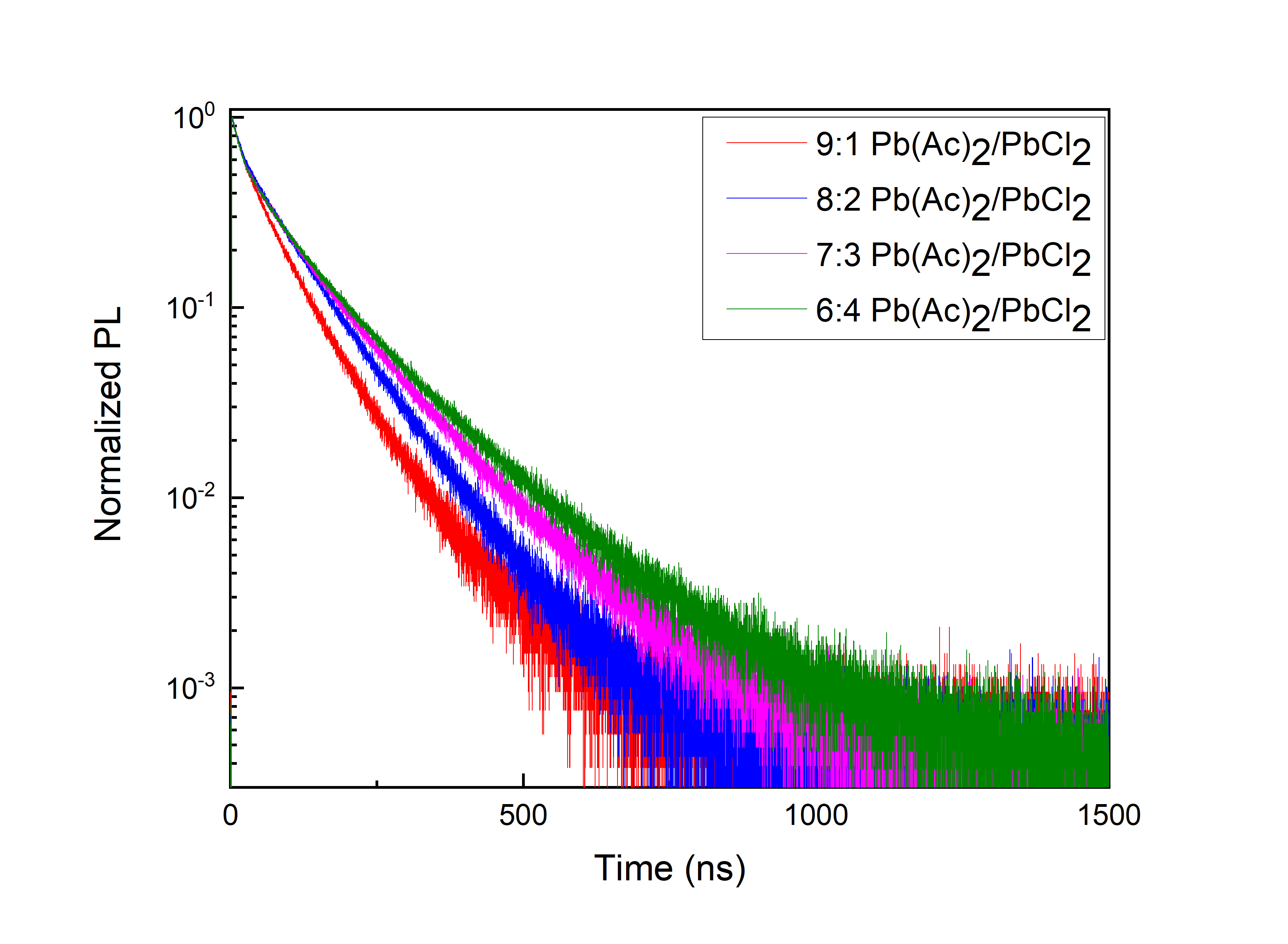}
  \caption{PL lifetimes of MAPbI$_3$ thin films with Pb(Ac)$_2$ / PbCl$_2$ ratios of 9:1, 8:2, 7:3 and 6:4}
  \label{fig:TRPL}
\end{figure}

Ferroelastic domain walls distort the local crystalline order and might form energetic barriers \cite{hermes2020anisotropic, coll2015polarization}. In our earlier study, we found an alternating polarization at the domain walls that could act as shallow traps states or scattering centres, affecting the charge carrier dynamics \cite{hermes2020anisotropic}. To study the effect of different domain morphologies on charge carrier lifetimes, we conducted time-resolved photoluminescence (TRPL) measurements on our MAPbI$_3$ film with different Pb(Ac)$_2$ / PbCl$_2$ ratios (Figure \ref{fig:TRPL}). We fitted the results bi-exponentially to obtain lifetime parameters. We found that the carrier lifetime increased from 68.47 ns for the film with 9:1 ratio to 104.35 ns for the 6:4 Pb(Ac)$_2$ / PbCl$_2$ ratio (all values are given in Table S1). This trend suggests the charged carriers move more freely in case of lower Pb(Ac)$_2$ / PbCl$_2$ ratios where the density of ferroelastic domain walls is decreased since the low energy barriers caused by the domain walls are eliminated. This result is in an agreement with our previous findings and previously proposed energetic barriers theory. Furthermore Zhang et al. discovered that the charge carrier mobility in MAPbI$_3$ is plane dependent i.e. anisotropic\cite{zhang2021carrier} which also offers an explanation for the energy barriers caused by ferroelastic twin domains.

To monitor the presence of additional phases and the optical properties, we performed UV-Vis spectroscopy on the MAPbI$_3$ films with all the different Pb(Ac)$_2$ / PbCl$_2$ ratios (Figure S11). All samples showed an absorption band edge around 780 nm, suggesting that the bandgap of MAPbI$_3$ did not change as a result of different precursor ratios. The absorption edge around 600 nm that is visible in all spectra comes from PbI$_2$ which is in an agreement with the PbI$_2$ signal that we observed in XRD measurements. Furthermore, at any Pb(Ac)$_2$ / PbCl$_2$ ratio below 8:2 a new peak at 390 nm emerged belonging to MAPbCl$_3$. This matches with the expectation since MAPbCl$_3$ forms when PbCl$_2$ is used or increased in the precursor solution \cite{medjahed2020unraveling}, which was also supported by ToF-SIMS results (Figure \ref{fig:vertical ToF-SIMS}) and non-annealed XRD results (S5, Supporting Information). Furthermore, the absorption decreased with decreasing Pb(Ac)$_2$ / PbCl$_2$ ratio due to decreasing amount of MAPbI$_3$ in the obtained film. This is possibly a result of the slow crystallization of MAPbI$_3$ due to the presence of PbCl$_2$. Nevertheless, the MAPbCl$_3$ phase could not be detected by the XRD measurements in Figure \ref{fig:XRD-All}, most likely due to the low MAPbCl$_3$ content in the films. In addition, we calculated Tauc plots (Figure S12, Supporting Information) from the absorption data  from UV-vis results (Figure S11, Supporting Information). The band gap of MAPbI$_3$ films slightly decreased with decreasing Pb(Ac)$_2$ / PbCl$_2$ ratio (Figure S13, Supporting Information). Since the bandgap of MAPbI$_3$ is the same for (110) and (001) planes \cite{zhang2021carrier}, the bandgap change can be attributed to increased strain as suggested elsewhere\cite{zhu2019strain}.

\section{Conclusion}

In this work we have shown a chemical route to control the strain in MAPbI$_3$ thin films by using different Pb(Ac)$_2$ / PbCl$_2$ ratios in the precursor solution. Using PFM and XRD, we monitored the internal structure, in particular the formation of ferroelastic twin domains - a clear signature of internal strain. The PFM results showed an increase in the average ferroelastic twin domain areas as the Pb(Ac)$_2$ / PbCl$_2$ ratio decreased. The XRD results confirmed that this increase was accompanied by an increase in lattice strain. We suggest that this strain increase is caused by the volatility difference of MA(Ac) and MACl which eventually leads to crystallization rate differences in various regions in the film. This difference leads to room temperature formation of highly strained MAPbI$_3$ in MA(Ac) areas and I$^-$ / Cl$^-$ exchange occurs in Cl-rich areas resulting in increased strain with increased PbCl$_2$ amount. The chemical gradient and depending strain gradient were observed via ToF-SIMS and XRD measurements, respectively.  To observe the effect of the resulting ferroelastic twin domain configurations on the charge carrier dynamics, we conducted TRPL measurements. The presence of larger ferroelastic twin domain areas i.e. fewer domain walls resulted in longer charge carrier lifetimes, supporting our earlier hypothesis of shallow energy barriers at twin domain walls\cite{hermes2020anisotropic}.

\

Our work shows that the crystal strain and thereby the formation of ferroelastic twin domains can be chemically manipulated by simply changing the ratio of different precursors without the need to apply any external force. This strategy offers a simple and scalable route to achieve a favorable domain arrangement in MAPbI$_3$-based devices, enabling faster and more efficient charge extraction. Even beyond domain engineering in MAPbI$_3$, chemical strain engineering is a promising idea for strain engineering in many other other lead halide perovskite materials.

\section{Experimental Section}
\subsection{Materials}

Methylalmmonium iodide (MAI) was purchased from Greatcell Energy. Lead acetate (Pb(Ac)$_2$ for perovskite precursor, 98.0 \% purity) and lead chloride (PbCl$_2$, 99.999 \%) were purchased from Tokyo Chemical Industry and Sigma Aldrich, respectively. Anhydrous dimethylformamide (DMF) was purchased from Sigma Aldrich.

\subsection{Perovskite Film Preparation}

\

Glass substrates were brushed with Hellmanex on both sides and washed with hot tap water and milliQ water followed by drying with an air gun. The cleaned substrates were subjected to UV-Ozone treatment (FHR UVOH 150 LAB, 250 W) for 30 minutes with oxygen feeding rate of 1 L/ min right before spin coating.

\

Two precursor solutions were prepared for the film preparation. Precursor solution 1 was prepared with MAI (477.0 mg, 3 mol), Pb(Ac)$_2$ (325.3 mg, 1 mol), and DMF (1 mL) while precursor solution 2 was prepared with MAI (238.5 mg, 1.5 mol), PbCl$_2$ (137.0 mg, 0.5 mol), and DMF (500 µL). The precursor solutions 1 and 2 were mixed in 9:1, 8:2, 7:3, and 6:4 volume ratios before the fabrication. The perovskite film fabrication was carried out by static spin coating, which was performed at 4000 rpms (1000 rpm/s)  for 1 minute after smearing 100 µL solution over the substrate. After the spin coating, all the samples were dried at room temperature for 10 minutes. After drying, all samples were annealed in DMF vapour atmosphere at 100 °C for 10 minutes. Glass petri dishes were used to obtain the DMF vapour atmosphere. The DMF vapour media were prepared before the coating started. The films first put on hotplate outside of the petri dish until it turns black ($\sim$ 3 seconds) then put under DMF vapour atmosphere. Number of samples per petri dish can be changed from 1 to 4. However, the amount of DMF should be changed accordingly. 15 µL DMF per film was used for the annealing process. All processes were carried out in a nitrogen glovebox.

\subsection{Piezoresponse Force Microscopy Measurements}

Piezoresponse force microscopy (PFM) is an Atomic force microscopy (AFM) mode, which is used to detect electromechanical material properties on the nanometer scale. A conductive tip is used to detect the response of the sample to an AC voltage.

\

Lateral piezoresponse force microscopy (PFM) measurements were performed on an Asylum Research MFP-3D AFM from Oxford Instruments and a Zürich Instruments HF2 lock-in amplifier in an argon filled glovebox (O$_2$ and H$_2$O $<$ 1ppm). The measurements were carried out with platinum-iridium coated SCM PIT-V2 cantilevers from Bruker with a nominal resonance frequency of 70 kHz and a spring constant of around 2.5 nN nm$^{-1}$. The measurement of the PFM signal was performed at an AC excitation voltage with a peak amplitude of 2 V at the lateral contact resonance at around 700 kHz to take advantage of the resonance enhancement. 

\subsection{X-ray Diffraction Measurements}

X-ray diffraction (XRD) patterns were taken with Rigaku SmartLab using Cu K$\alpha$ radiation. Diffractograms had been recorded in “$\theta$”- “$\theta$” geometry, using a Göbel mirror and automatic sample alignment. The scanning rate was 2\textdegree/min in steps of 0.01\textdegree{}. The total beam exposure time was 30 minutes.

\subsection{Time of Flight Secondary Ion Mass Spectroscopy}

ToF-SIMS experiments were performed using a TOF.SIMS5 (NCS) instrument (IONTOF GmbH, Münster, Germany) with 30 keV Bi3 primary ions and 5 keV Ar1500 cluster ions for sputtering at a 45° angle. Surface imaging was facilitated using the ultimate imaging mode at a current of 0.05 pA at a cycle time of 100 µs. Dual beam depth profiling was conducted on an analysis area of 200 × 200 µm$^2$ with a sputter area of 500 x 500 µm$^2$ at a sputter current of 2.5 nA.

\subsection{Optical Measurements}

Absorption data was collected using a Cary5000 UV-vis-NIR spectrometer by Agilent equipped with an integrating sphere. Time-resolved photoluminescence was conducted via time-correlated single photon counting on a FluoTime300 spectrometer by PicoQuant at an emission wavelength of 770nm, following pulsed 485nm excitation at a repetition rate of 500kHz.

\medskip
\textbf{Acknowledgements} \par 
Y.Y. and S.A.L.W. acknowledge the SPP2196 project (Deutsche Forschungsgemeinschaft) for funding. K.A.-W. acknowledges support from the Carl-Zeiss-Stiftung.

\medskip

\clearpage
\bibliographystyle{unsrt}
\bibliography{References}


\clearpage
\title{Chemical Strain Engineering of MAPbI$_3$ Perovskite Films \\ Supporting Information}
\maketitle

\renewcommand{\thepage}{S\arabic{page}} 
\renewcommand{\thesection}{S\arabic{section}}  
\renewcommand{\thetable}{S\arabic{table}}  
\renewcommand{\thefigure}{S\arabic{figure}}
\setcounter{figure}{0}  
\setcounter{section}{0}
\setcounter{page}{1}

\section*{Piezoresponse Force Microscopy}

\begin{figure*}[!htb]
    \begin{center}
       \includegraphics[width=\linewidth]{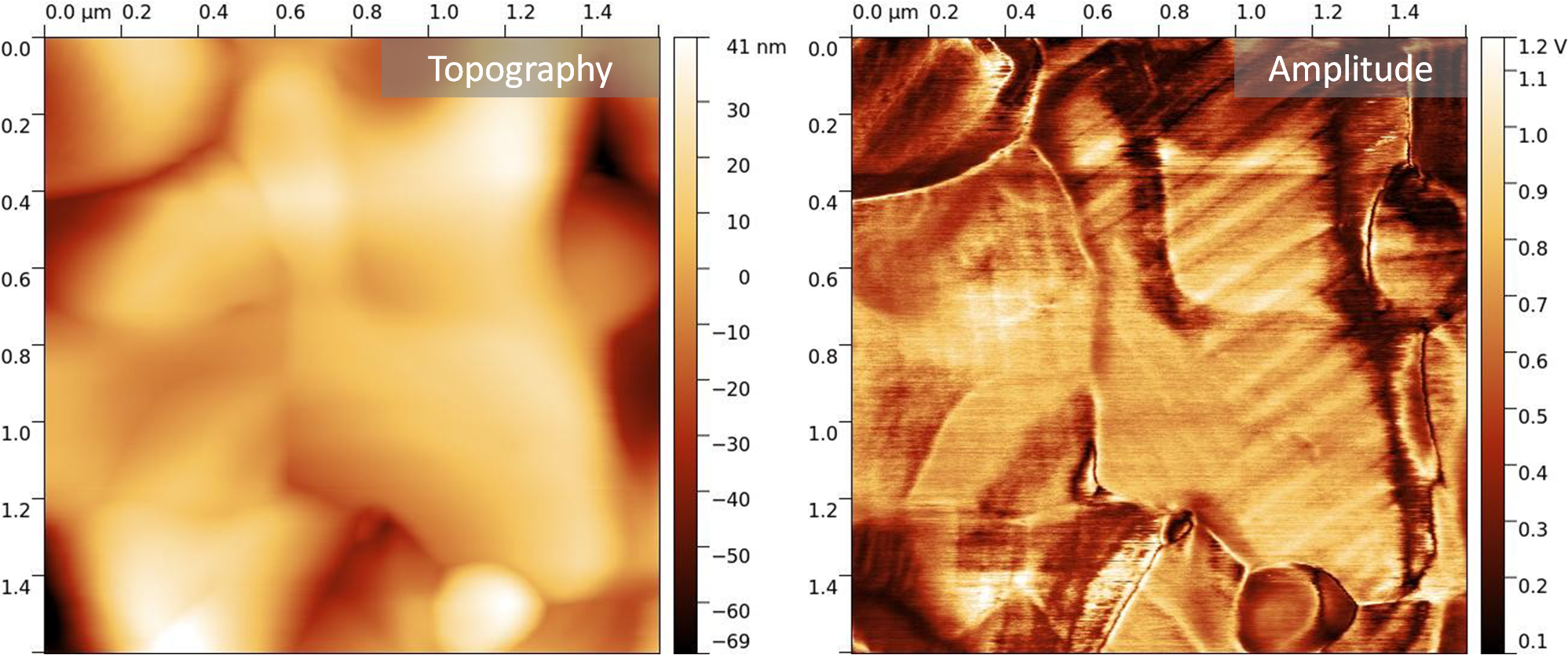}
      
       \caption{Topography and lateral PFM images of MAPbI$_3$ film made with only MAI and Pb(Ac)$_2$}
    \end{center}
\end{figure*}

\begin{figure*}[!htb]
    \begin{center}
       \includegraphics[width=\linewidth]{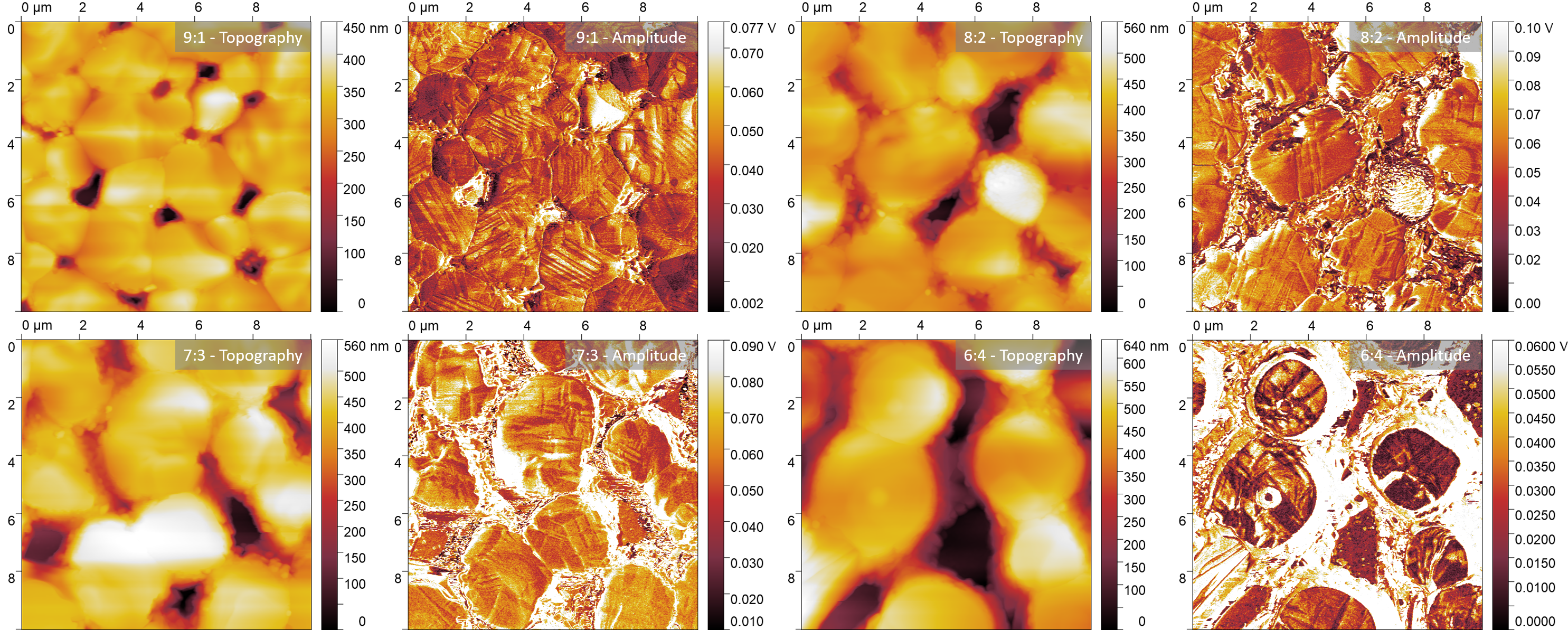}
      
       \caption{Topography and lateral PFM images of MAPbI$_3$ films with Pb(Ac)$_2$ / PbCl$_2$ ratios of 9:1, 8:2, 7:3 and 6:4}
    \end{center}
\end{figure*}

\clearpage

\section*{Bragg's Law}
The Bragg's law gives the relation between the diffraction angle and the interplanar distance as follows:

\begin{equation}
    n\lambda = 2d sin\theta
\end{equation}

where \textlambda, d, and \straighttheta \ denotes interplanar distance, and diffraction angles, respectively. Furthermore, the interplanar distance for the tetragonal lattice is given as

\begin{equation}
    \frac{1}{d^{2}} = \frac{(h^{2}+k^{2})}{a^{2}} + \frac{l^{2}}{c^{2}}
\end{equation}

where h,k, and l are the Miller indices and a and c are lattice parameters. When Equation S1 and S2 are combined, it can be seen that the reflection angles for {hkl} planes in XRD may give information on the lattice parameters. A shift to higher scattering angles would indicate lattice shrinkage whereas a shift to lower scattering angles would indicate lattice expansion. Therefore, the peak shift to lower scattering angles in Figure 2 can be interpreted as possible lattice expansion.

\section*{X-ray Diffraction Measurements}

\begin{figure*}[h!]
    \begin{center}
       \includegraphics[width=0.5\linewidth]{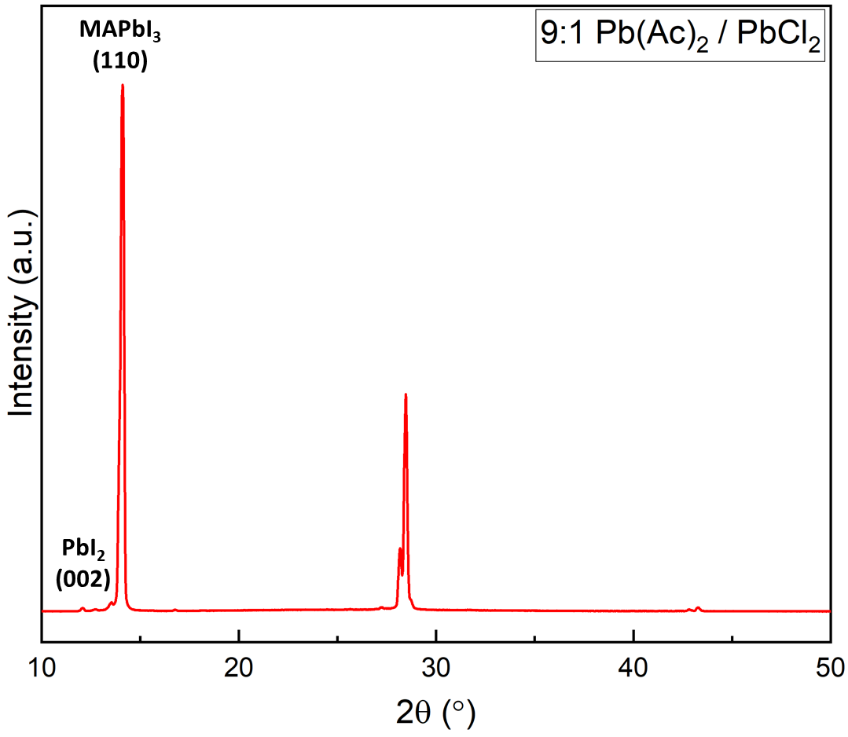}
       \caption{XRD pattern of MAPbI$_3$ film made from 9:1 Pb(Ac)$_2$ / PbCl$_2$ ratio.}
    \end{center}
\end{figure*}

\begin{figure*}[h!]
    \begin{center}
       \includegraphics[width=\linewidth]{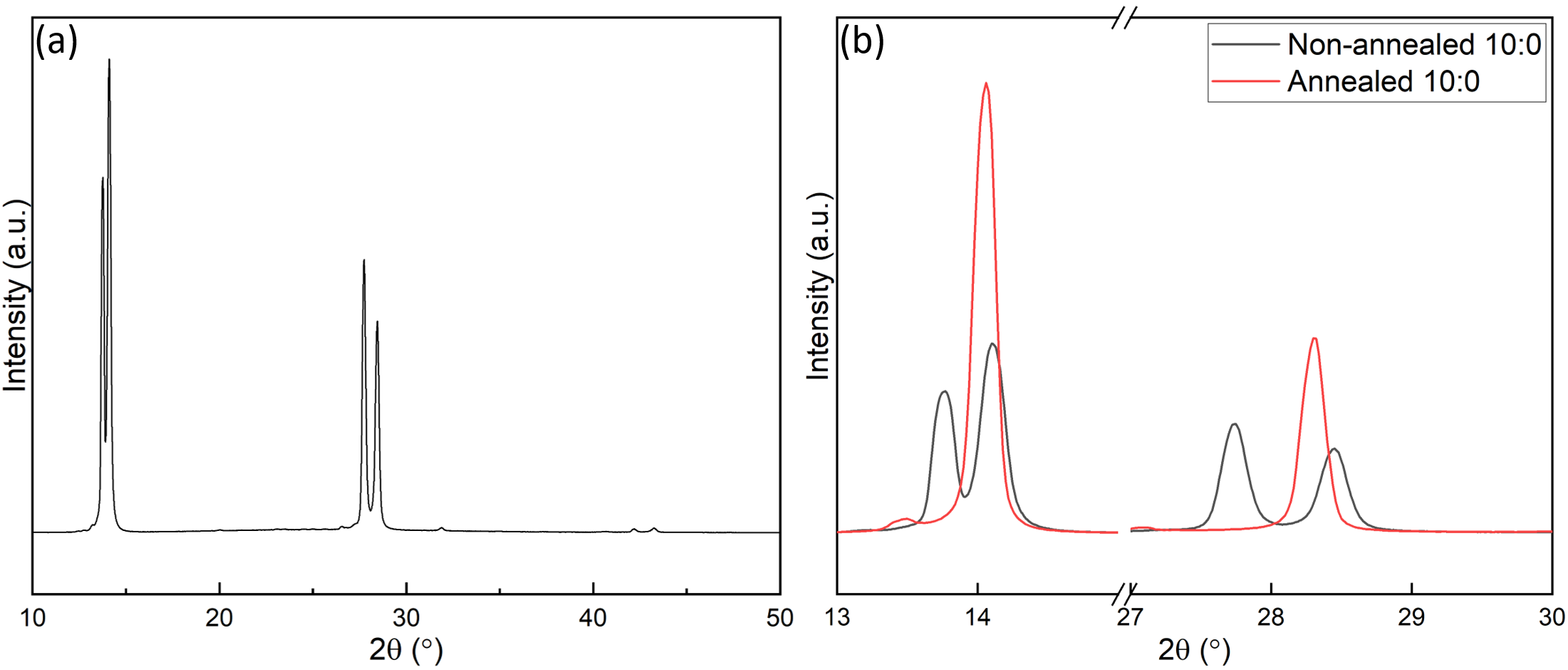}
       \caption{XRD pattern of MAPbI$_3$ film made with only MAI and Pb(Ac)$_2$ (a) without and (b) with annealing}
    \end{center}
\end{figure*}

\begin{figure*}[b]
    \begin{center}
       \includegraphics[width=\linewidth]{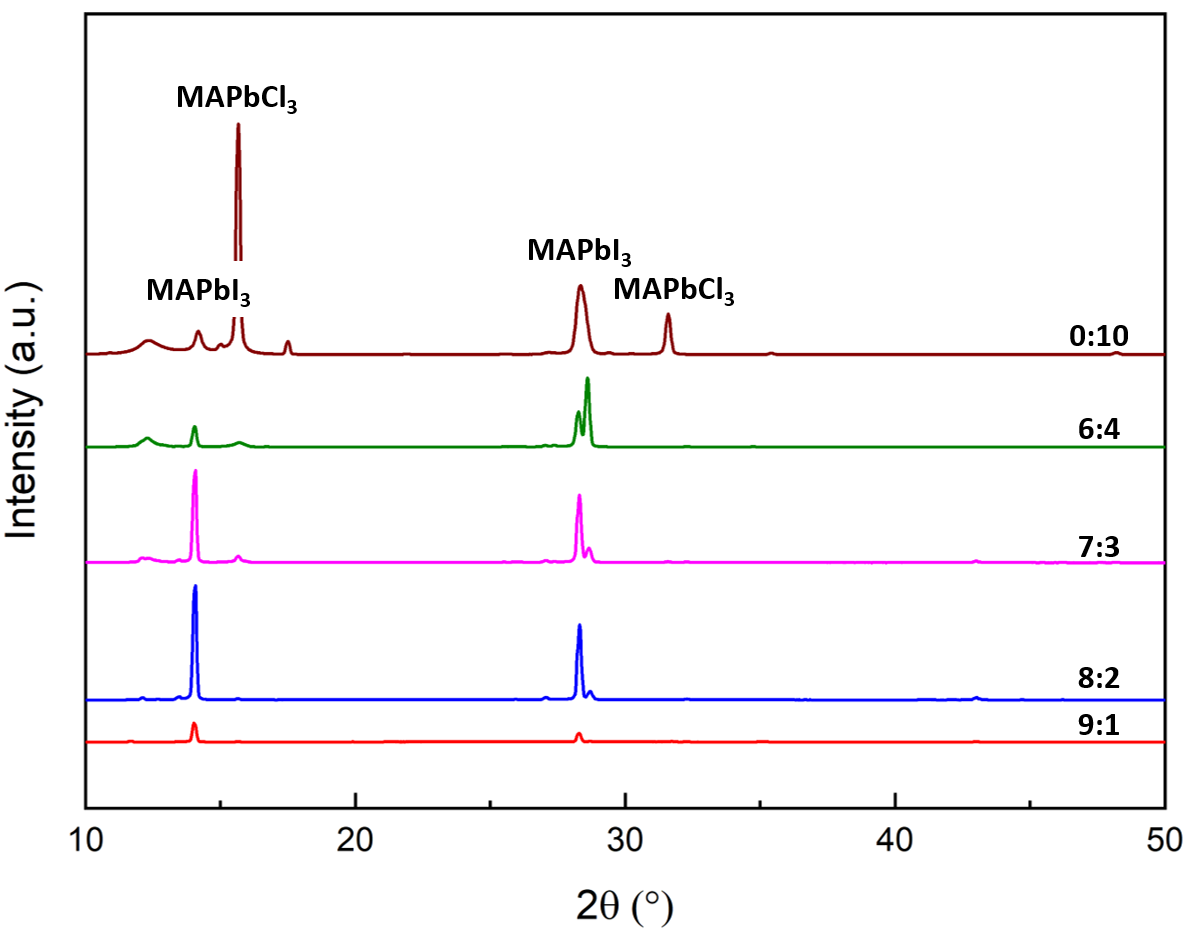}
       \caption{XRD pattern of non-annealed MAPbI$_3$ films with Pb(Ac)$_2$ / PbCl$_2$ ratios of 9:1, 8:2, 7:3, 6:4, and 0:10}
       \label{fig:Non-Annealed XRD}
    \end{center}
\end{figure*}

In order to gain insight on the twin formation, we carried out XRD measurements on MAPbI$_3$ films with different precursor ratios (Figure \ref{fig:Non-Annealed XRD}, Supporting Information) right after the spin coating without annealing. First notable observation here is the increasing intensity of MAPbCl$_3$ reflection upon increasing the PbCl$_2$ content in the precursor, which supports the I$^-$ / Cl$^-$ ion exchange theory. Even though the MAPbCl$_3$ reflection is present in all spectra in Figure S6, the intensity of the peaks is extremely low compared to a pure MAI/PbCl$_2$ film. This could be caused by the formation of a Cl-rich perovskite region at the interface to the substrate. Another important observation is the absence of twinning in MAI/PbCl$_2$ film while all other samples show (004)/(220) twinning. These results suggest the room temperature crystallization of tetragonal MAPbI$_3$ due to high volatility of MA(Ac) leads to high strain - ergo twinning. On the other hand, in case of twinning in MAPbI$_3$ film made from MAI/PbCl$_2$ precursors, the I$^-$ / Cl$^-$ ion exchange is the main reason for the strain. Therefore, we speculate that the increased strain in MAPbI$_3$ films from Pb(Ac)$_2$ / PbCl$_2$ precursors is the combined result of having highly strained MAPbI$_3$ grains in Pb(Ac)$_2$-rich areas due to room temperature crystallization and I$^-$ / Cl$^-$ ion exchange in PbCl$_2$-rich areas during MAPbI$_3$ formation.

\clearpage

\section*{Grazing Incidence X-ray Diffraction}

\begin{figure*}[h!]
    \begin{center}
       \includegraphics[width=\linewidth]{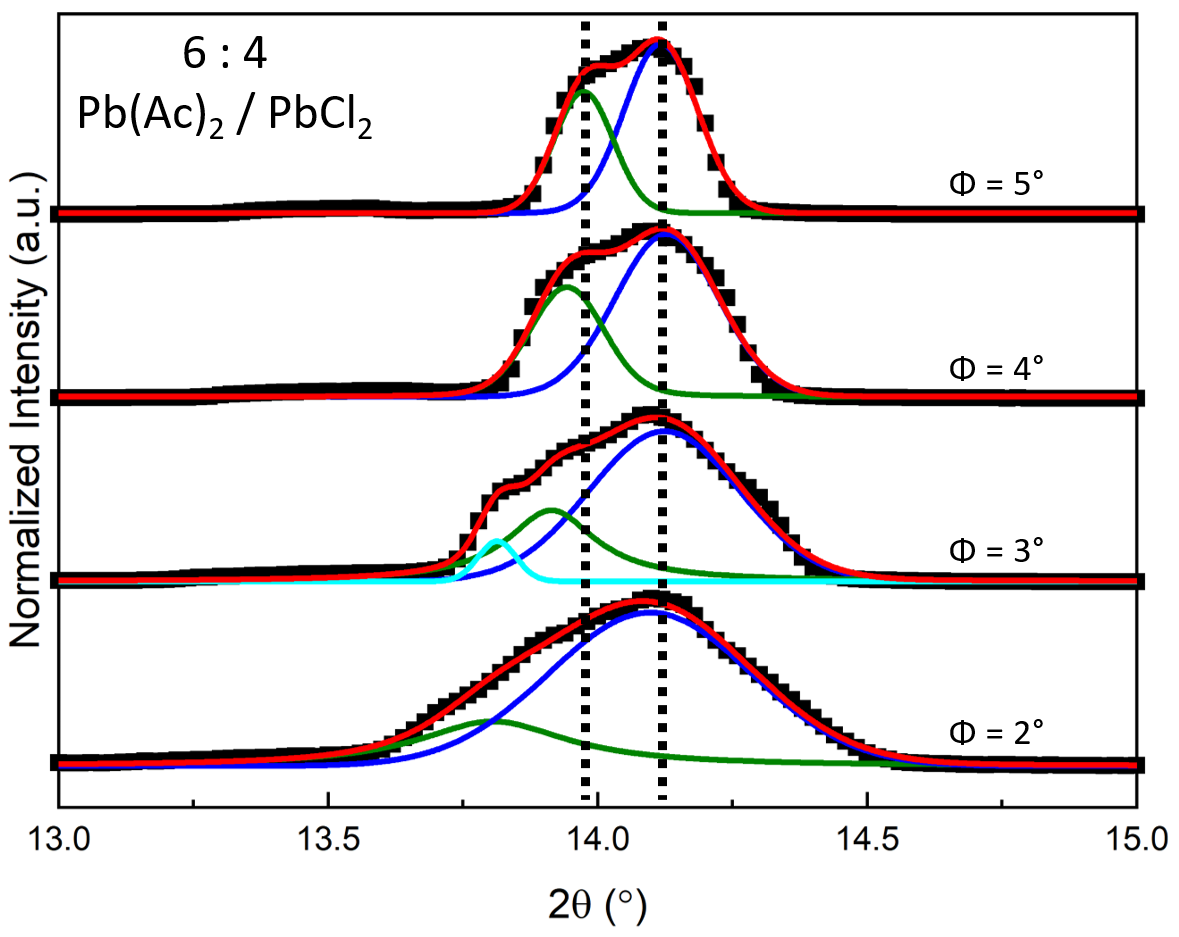}
       \caption{Gracing incidince angle dependent XRD reflections of MAPbI$_3$ film made with 6:4 Pb(Ac)$_2$ / PbCl$_2$ ratio}
    \end{center}
\end{figure*}

\newpage
\pagebreak  [4]

\section*{Time-of-Flight Secondary Ion Mass Spectroscopy}

\begin{figure*}[h!]
    \begin{center}
       \includegraphics[width=\linewidth]{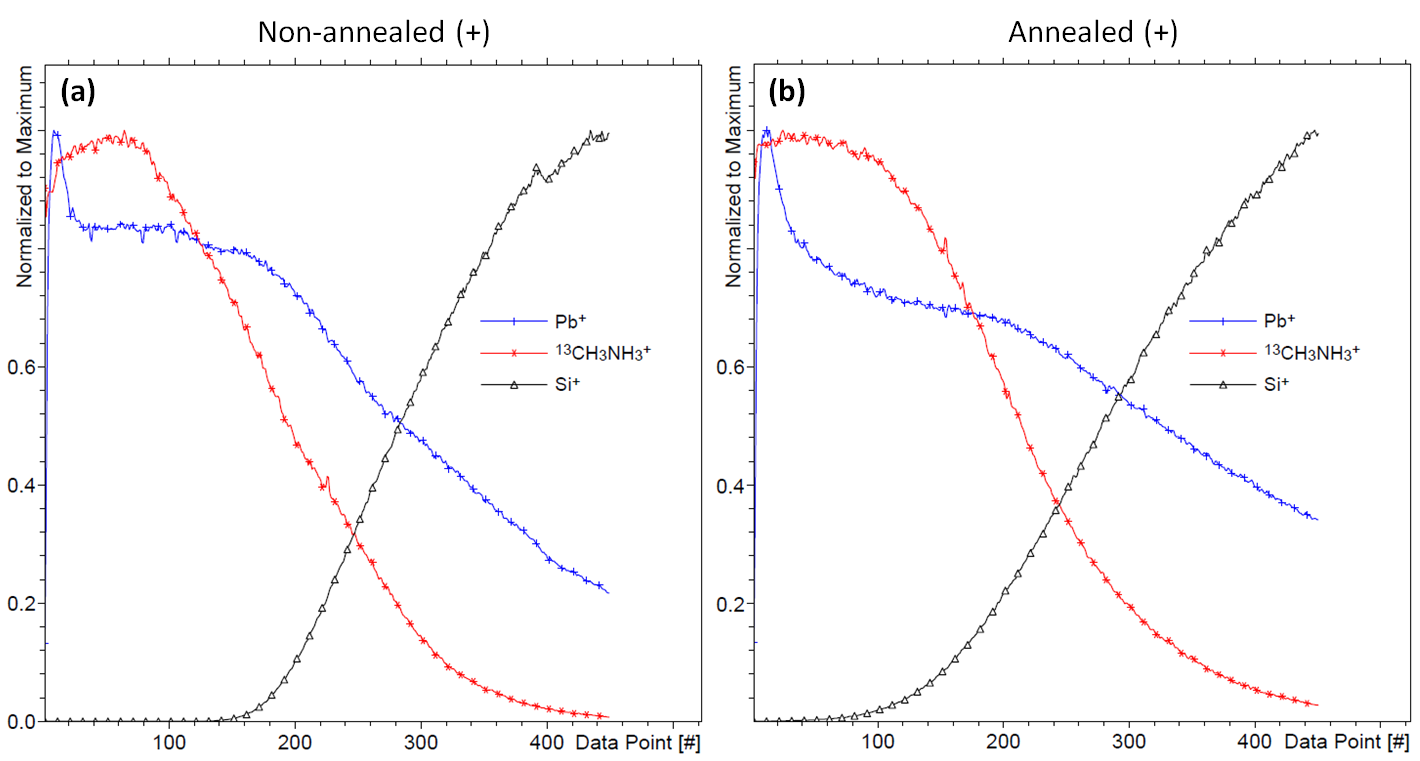}
       \caption{Positive chemical depth profiles of (a) non-annealed and (b) annealed MAPbI$_3$ film made with 6:4 Pb(Ac)$_2$ / PbCl$_2$ ratio.}
    \end{center}
\end{figure*}

\begin{figure*}[h!]
    \begin{center}
       \includegraphics[width=\linewidth]{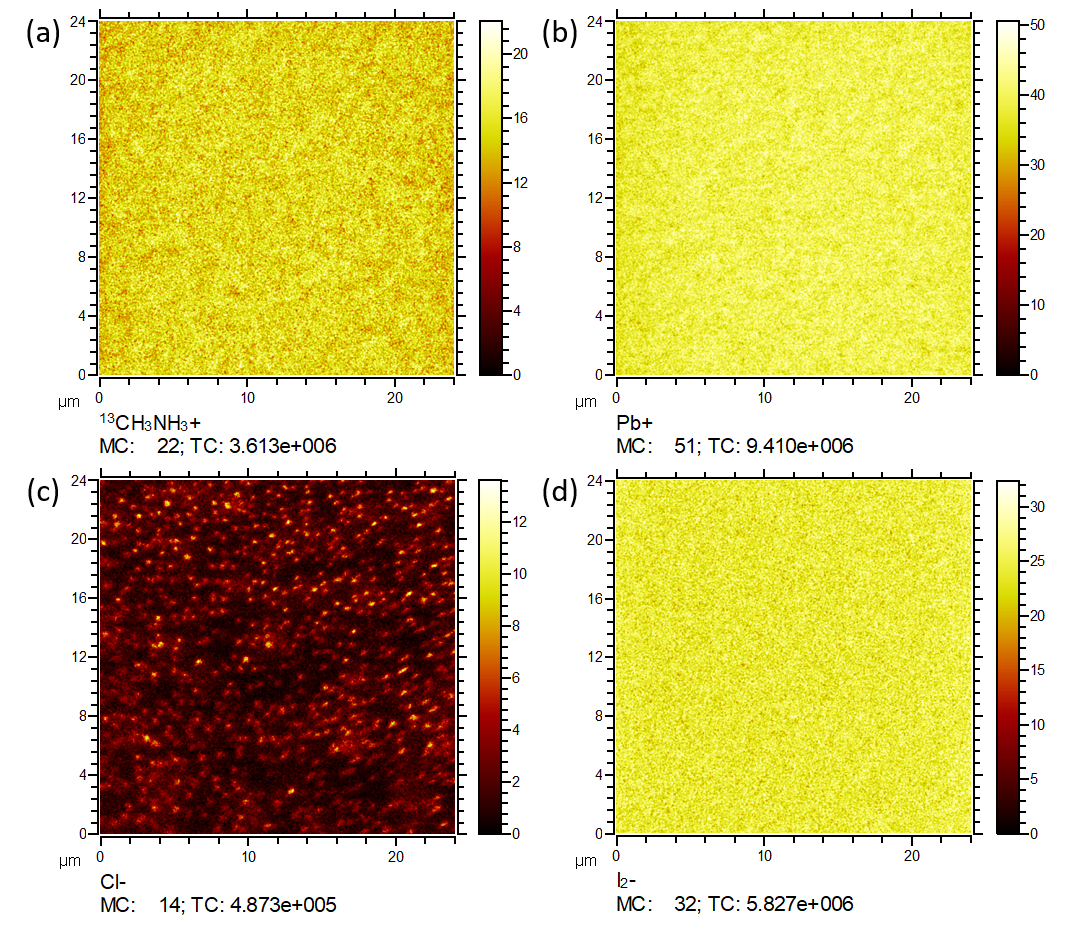}
       \caption{Lateral chemical cation distributions in non-annealed MAPbI$_3$ film made with 6:4 Pb(Ac)$_2$ / PbCl$_2$ ratio, 24 µm x 24 µm}
    \end{center}
\end{figure*}

\begin{figure*}[h!]
    \begin{center}
       \includegraphics[width=\linewidth]{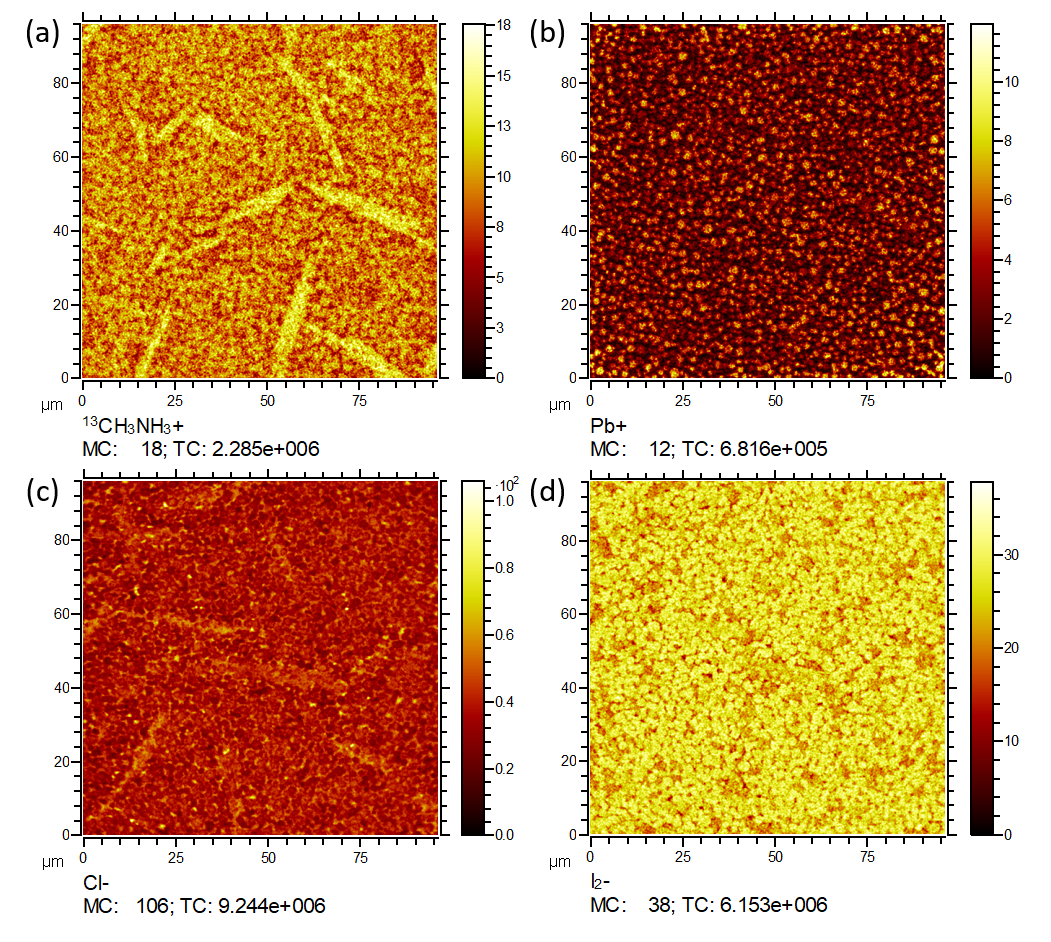}
       \caption{Lateral chemical ion distributions in annealed MAPbI$_3$ film made with 6:4 Pb(Ac)$_2$ / PbCl$_2$ ratio, 100 µm x 100 µm}
    \end{center}
\end{figure*}

\begin{figure*}[h!]
    \begin{center}
       \includegraphics[width=\linewidth]{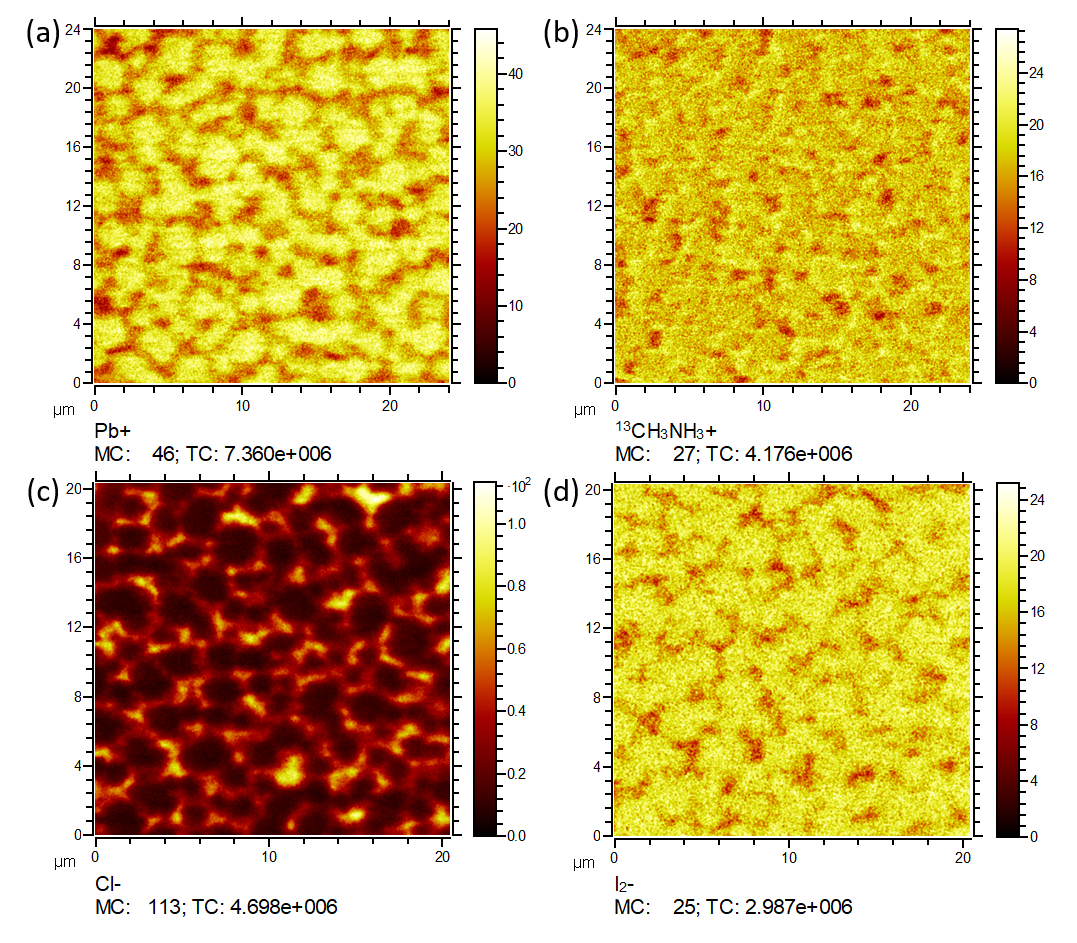}
       \caption{Lateral chemical ion distributions in annealed MAPbI$_3$ film made with 6:4 Pb(Ac)$_2$ / PbCl$_2$ ratio, 24 µm x 24 µm}
    \end{center}
\end{figure*}

\clearpage

\section*{Bandgap and Time-resolved Photoluminescence Fit Results}

\begin{figure*}[h!]
    \begin{center}
       \includegraphics[width=\linewidth]{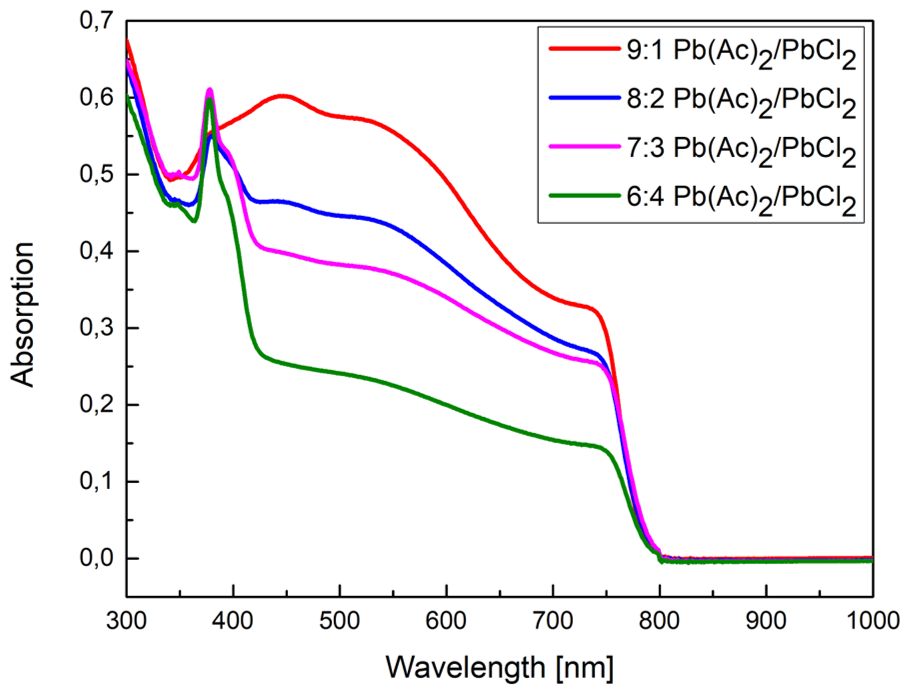}
      
       \caption{UV-Vis measurement results of MAPbI$_3$ thin films with Pb(Ac)$_2$ / PbCl$_2$ ratios of (a) 9:1, (b) 8:2, (c) 7:3 and (d) 6:4}
    \end{center}
\end{figure*}

\begin{figure*}[h!]
    \begin{center}
       \includegraphics[width=\linewidth]{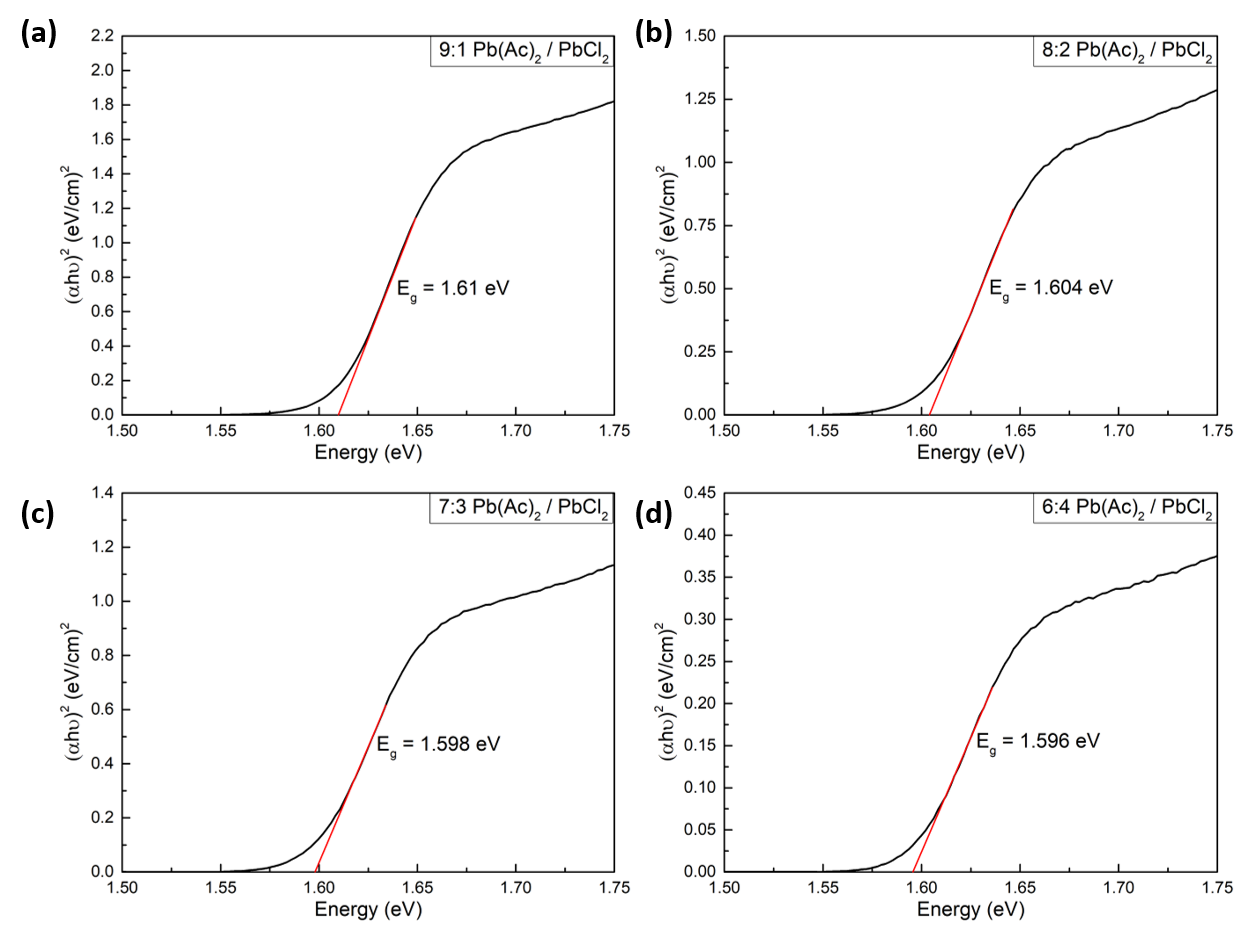}
      
       \caption{Tauc plots of MAPbI$_3$ thin films with Pb(Ac)$_2$ / PbCl$_2$ ratios of (a) 9:1, (b) 8:2, (c) 7:3 and (d) 6:4}
    \end{center}
\end{figure*}

\begin{figure*}[h!]
    \begin{center}
       \includegraphics[width=\linewidth]{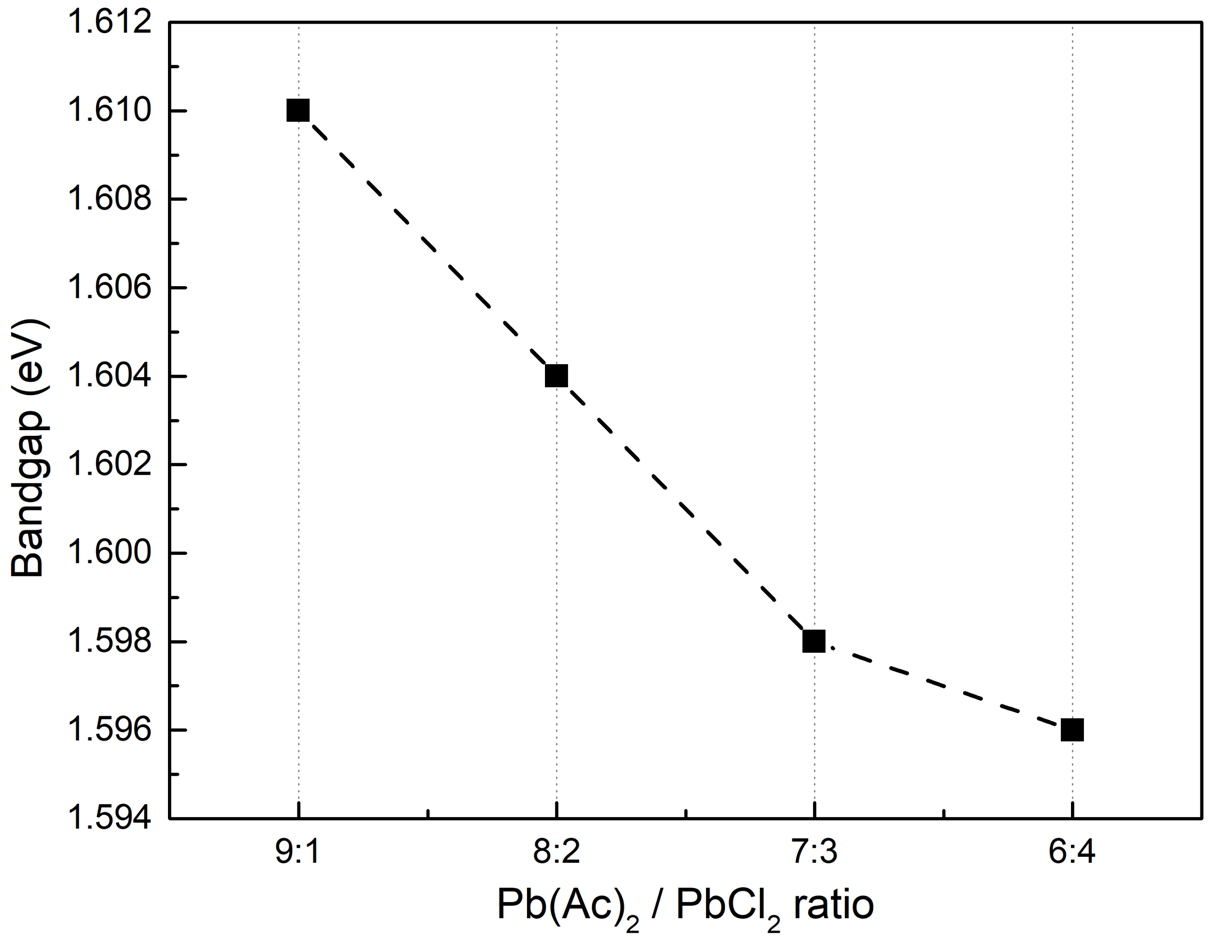}
      
       \caption{Pb(Ac)$_2$ / PbCl$_2$ ratio dependent bandgaps of MAPbI$_3$ films}
    \end{center}
\end{figure*}

\clearpage

\begin{table*}[h!]
  \caption{TRPL fit results for MAPbI$_3$ thin films with Pb(Ac)$_2$ / PbCl$_2$ ratios of 10:0, 9:1, 8:2, 7:3 and 6:4}
  \label{tbl:excel-table}
  \includegraphics[width=\linewidth]{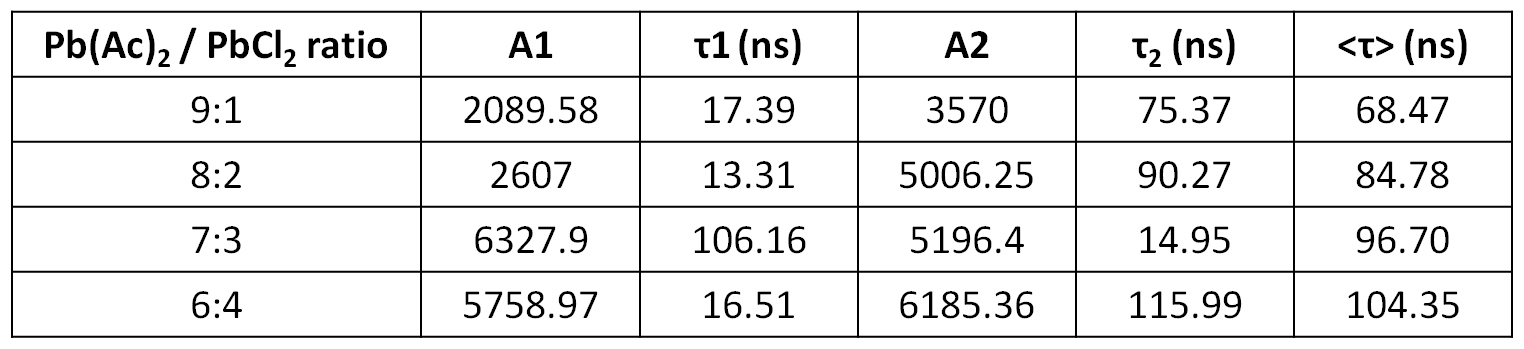}
\end{table*}

\section*{SEM Images}

\begin{figure*}[h!]
    \begin{center}
       \includegraphics[width=\linewidth]{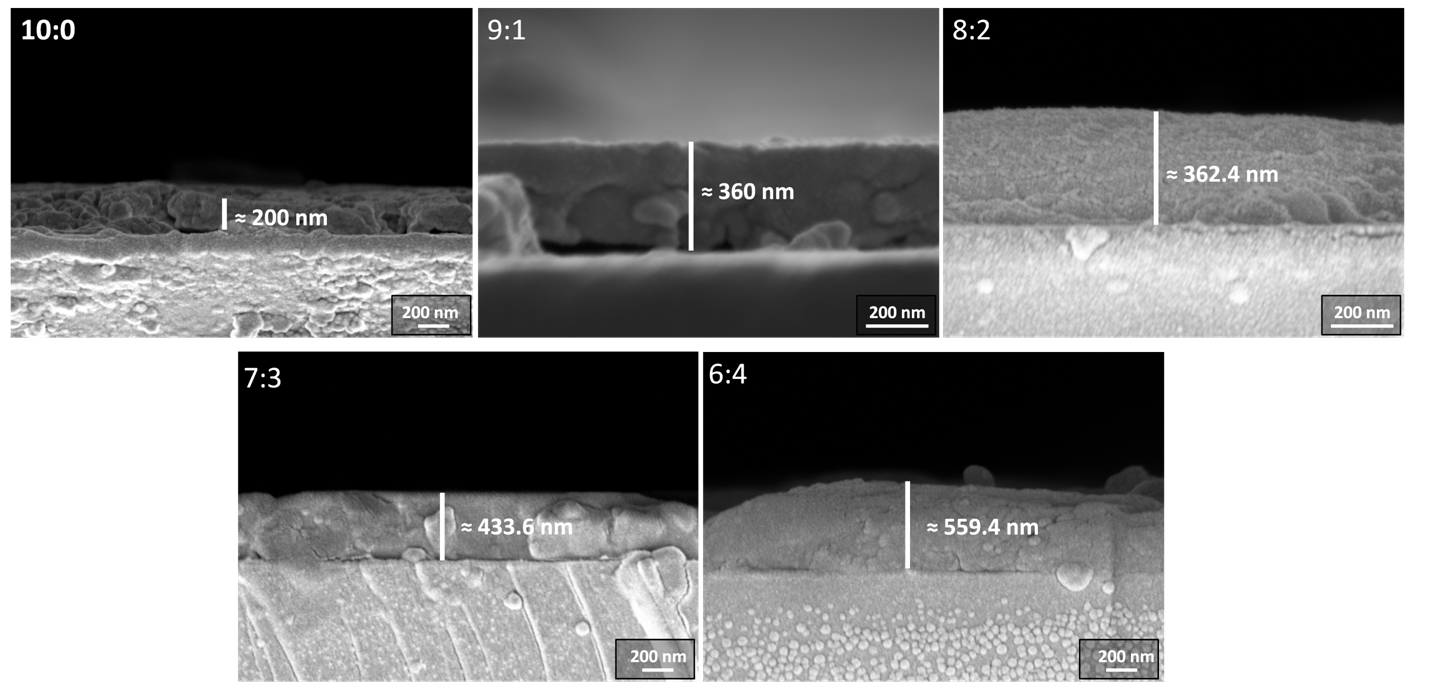}
      
       \caption{SEM images of MAPbI$_3$ films with different Pb(Ac)$_2$ / PbCl$_2$ ratio.}
    \end{center}
\end{figure*}

\end{document}